\documentclass[twocolumn,showkeys,aps,prb,showpacs]{revtex4-1}
\usepackage{graphicx}
\usepackage[CJKbookmarks,dvipdfm,colorlinks,linkcolor=blue,citecolor=blue]{hyperref}

\begin{document}

\title{Isolated highly localized bands in $\mathrm{YbI_2}$ monolayer caused by $4f$ orbitals}

\author{San-Dong Guo}
\affiliation{Department of Physics, School of Sciences, China University of Mining and
Technology, Xuzhou 221116, Jiangsu, China}
\begin{abstract}
The novel electronic structures  can induce  unique physical properties in  two-dimensional (2D) materials. In this work, we report isolated highly localized bands in $\mathrm{YbI_2}$ monolayer by the first-principle calculations within generalized gradient approximation (GGA) plus spin-orbit coupling (SOC). It is found that $\mathrm{YbI_2}$ monolayer is an indirect-gap semiconductor using both GGA and GGA+SOC.  The calculations reveal that Yb-$4f$ orbitals constitute isolated highly localized bands below the Fermi level at the absence of SOC, and the bands are split into the $j = 7/2$  and $j = 5/2$  states with SOC. The isolated highly localized bands can lead to very large Seebeck coefficient and very low electrical conductivity in p-type doping by producing very large effective mass of the carrier. It is proved that isolated highly localized bands have very strong stability again strain, which is very important for practical application. When the onsite Coulomb interaction  is added to the Yb-$4f$ orbitals, isolated highly localized bands persist, and only their  relative positions in the gap change. These findings open a new window to search for novel electronic structures in 2D materials.

\end{abstract}
\keywords{Monolayer; Spin-orbit coupling; Novel electronic structures; Strain}

\pacs{71.20.-b, 73.22.-f, 72.20.-i, 74.62.F ~~~~~~~~~~~~~~~~~~~~~~~~~~~~~~~~~~~Email:sandongyuwang@163.com}

\maketitle

\section{Introduction}
Since the discovery of graphene\cite{q6}, 2D materials have attracted enormous research
interest in electronic, optical, topological and thermal properties.  A large amount of 2D materials have been predicted theoretically, or achieved experimentally, such as transition metal dichalcogenide (TMD), group-VA, group IV-VI and group-IV  monolayers\cite{q7,q8,q9,q10,q11}.
Graphene has a peculiar electronic structure with the dispersion relation being  linear around the Fermi level, and the related electrons and holes need be described by the Dirac equation\cite{q6}.  Compared with the gapless graphene,  $\mathrm{MoS_2}$ as a representative semiconducting TMD monolayer has triggered  a new wave of research in TMD monolayers due to potential application for novel ultrathin and flexible devices\cite{q12,q13}. Recently,  Janus monolayer MoSSe has been  experimentally synthesized  by  replacing the top S atomic layer of  $\mathrm{MoS_2}$  with Se atoms\cite{q14}, which provides more possibilities for
more extensive nanoelectronic and optoelectronic applications. Phosphorene,  possessing novel high carrier mobility and intrinsically large fundamental direct band gap,  has great prospective for its applications  in field-effect transistors and photo-transistors\cite{q15,q16}.
Experimentally, the 2D Dirac nodal line fermions has been reported  in monolayer $\mathrm{Cu_2Si}$, which provides a platform to study the novel physical properties in 2D Dirac materials\cite{q17}. Thermal transports of 2D materials have been widely investigated, including  external perturbation like strain, substrate  and clustering\cite{q18}.

\begin{figure}
  \includegraphics[width=7.0cm]{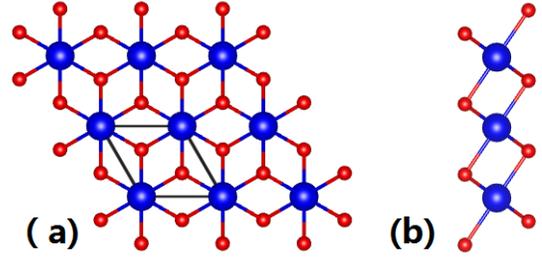}
  \caption{(Color online) The schematic crystal structure of $\mathrm{YbI_2}$ monolayer. The blue balls represent Yb atoms, and the red  balls for I atoms.}\label{t0}
\end{figure}
\begin{figure}
  \includegraphics[width=8cm]{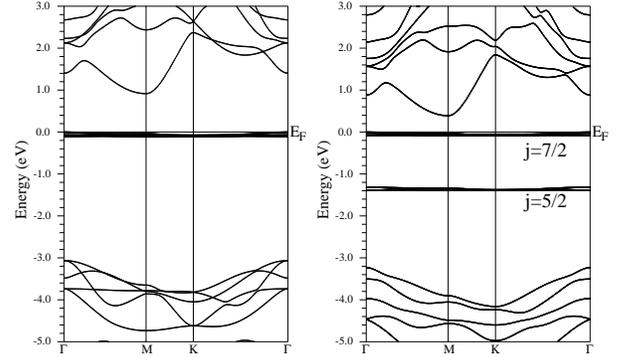}
\caption{The energy band structures  of $\mathrm{YbI_2}$ monolayer  using GGA (Left) and GGA+SOC (Right). }\label{t1}
\end{figure}

\begin{figure*}
  \includegraphics[width=12cm]{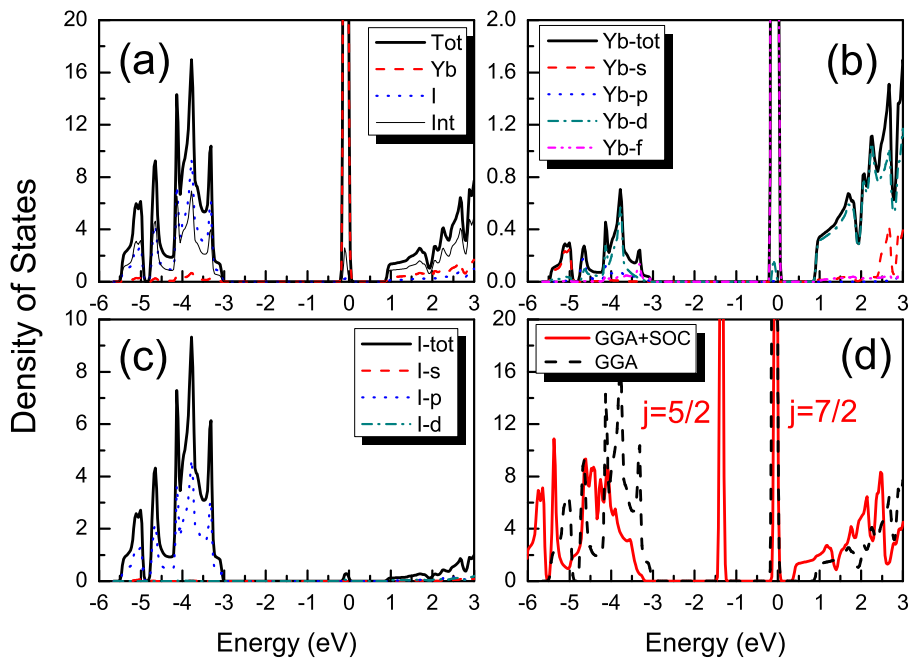}
\caption{(a,b,c) the total and projected DOS  of  $\mathrm{YbI_2}$ monolayer using GGA; (d) the total DOS using GGA and GGA+SOC. }\label{t2}
\end{figure*}

\begin{figure}
  \includegraphics[width=8cm]{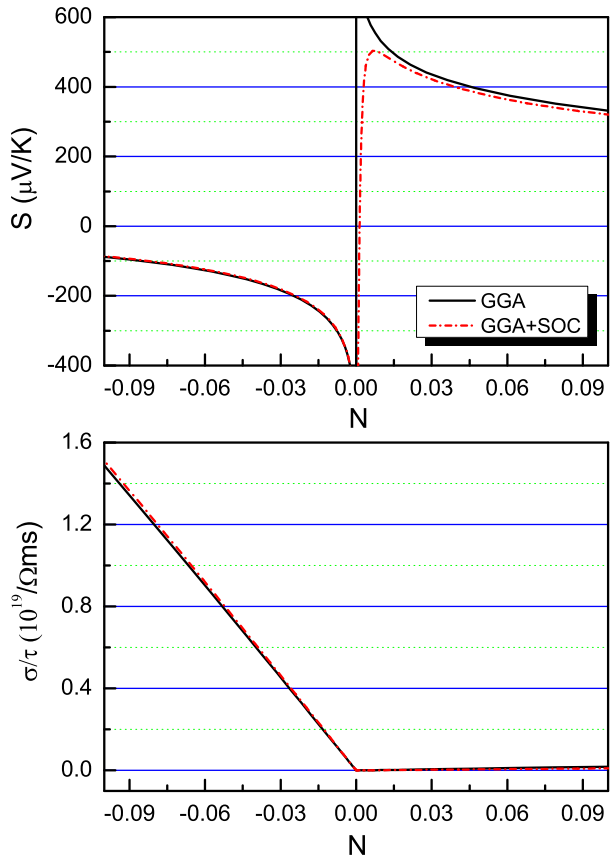}
\caption{At 300 K, the transport coefficients of $\mathrm{YbI_2}$ monolayer as a function of doping level (N) using GGA and GGA+SOC: Seebeck coefficient S and electrical conductivity with respect to scattering time  $\mathrm{\sigma/\tau}$. }\label{t3}
\end{figure}

\begin{figure*}
  \includegraphics[width=12cm]{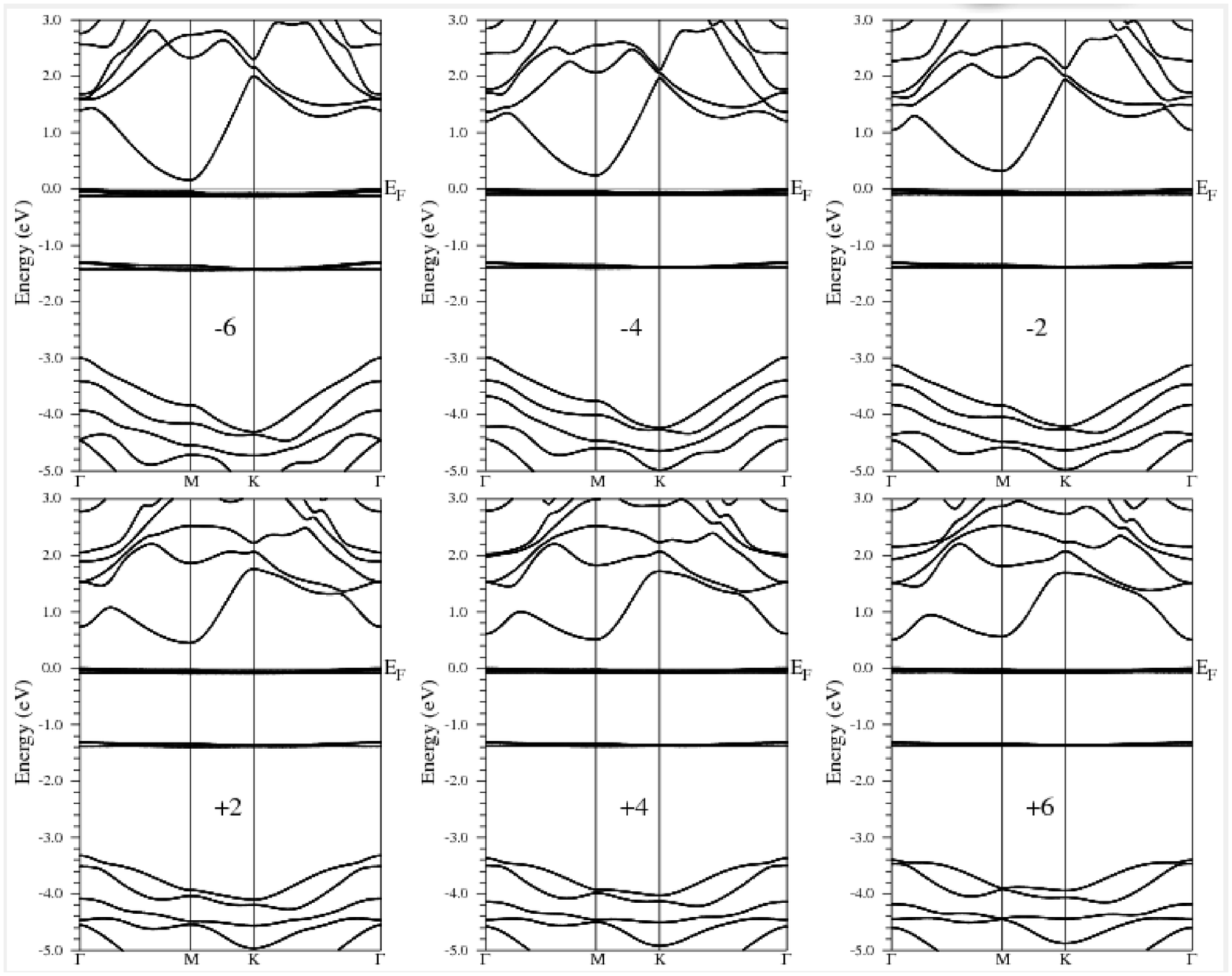}
  \caption{(Color online) The energy band structures  of $\mathrm{YbI_2}$ monolayer with $\varepsilon$ changing from -6\% to 6\%  using GGA+SOC.}\label{t4}
\end{figure*}

Searching for other peculiar electronic structure, like isolated highly localized bands,  would be of great significance to the the design and development of  nano-devices. Recently, the plentiful 2D materials are predicted from high-throughput
computational exfoliation of experimentally known compounds\cite{q19}. Among them, $\mathrm{YbI_2}$ monolayer with $4f$ electrons is predicted, which is interesting to investigate its electronic structure due to localized $4f$ orbitals.
In this work, the electronic structures of  $\mathrm{YbI_2}$  monolayer  are studied by  first-principles calculations.  It is found that there are some isolated highly localized bands with Yb-$4f$ character in a very large gap.  The SOC can  split the $4f$ bands  into   $j = 7/2$  and $j = 5/2$  states, and the splitting  gap is up to about 1.22 eV. In p-type doping, the very large Seebeck coefficient and very low electrical conductivity can be found due to very large effective mass of the  p-type carrier caused by the isolated highly localized bands. Calculated results show that isolated highly localized bands are very stable  again strain, and the electron correlation effects only change the relative positions of  highly localized bands in the gap.

\section{Computational detail}
Within the density functional theory (DFT) \cite{1}, we use a full-potential linearized augmented-plane-waves method to investigate
electronic structures  of   $\mathrm{YbI_2}$   monolayer, as implemented in
the WIEN2k  code\cite{2}.
The popular GGA of Perdew, Burke and  Ernzerhof  (GGA-PBE)\cite{pbe} is used as the
exchange-correlation potential, and the  internal atomic positions are optimized with a force standard of 2 mRy/a.u..
Due to containing heavy element Yb, the SOC was included self-consistently \cite{10,11,12,so}.
 To attain reliable results, we use  30$\times$30$\times$1 k-meshes in the
first Brillouin zone (BZ) for the self-consistent calculation with harmonic expansion up to $\mathrm{l_{max} =10}$  and  $\mathrm{R_{mt}*k_{max} = 8}$.
The self-consistent calculations are
considered to be converged when the integration of the absolute
charge-density difference between the input and output electron
density is less than $0.0001|e|$ per formula unit, where $e$ is
the electron charge.
Based on calculated energy band
structures, the  Seebeck coefficient and electrical conductivity of $\mathrm{YbI_2}$ are performed through solving Boltzmann
transport equations within the constant
scattering time approximation (CSTA),  as implemented in
BoltzTrap software\cite{b}. To achieve the convergence results, the parameter LPFAC is set to 40, and 100$\times$100$\times$1 k-meshes in the
first BZ is used  for the energy band calculation.

\section{MAIN CALCULATED RESULTS AND ANALYSIS}
\autoref{t0} shows the structure of $\mathrm{YbI_2}$ monolayer, containing three atomic sublayers with Yb layer sandwiched  I layers.
The similar structure can also be found in TMD monolayer\cite{q20}, such as  $\mathrm{ZrS_2}$ and  $\mathrm{PtSe_2}$ monolayers with 1T phase.
However, it is different from the  $\mathrm{MoS_2}$  as a representative TMD monolayer with 2H phase.
 The unit cell  of $\mathrm{YbI_2}$ monolayer  is built with the vacuum region of more than 18 $\mathrm{{\AA}}$ to avoid spurious interaction between neighboring layers.   The optimized lattice constants $a$ ($b$) using GGA  is 4.46 $\mathrm{{\AA}}$ with Yb and
I atoms occupying the (0, 0, 0) and  (1/3, 2/3, 0.922) positions, respectively. The Yb-I bond length is 3.14 $\mathrm{{\AA}}$, and I-Yb-I bond angle for $89.265^\circ$  ($90.735^\circ$), and  the thicknesses of   $\mathrm{YbI_2}$ monolayer for 3.58  $\mathrm{{\AA}}$.

The calculated energy band structures of $\mathrm{YbI_2}$ monolayer are shown in \autoref{t1}  with GGA and GGA+SOC, and  the related density of states (DOS) are plotted in \autoref{t2}. Without SOC,  fourteen highly localized bands with Yb-$4f$  character are observed  near the Fermi level, and the bandwidth is only 0.118 eV. When including SOC, the $4f$ bands are split into the $j = 7/2$  and $j = 5/2$  states, producing a gap of 1.218 eV. Similar splitting can also be found in  $\mathrm{YbB_6}$ by SOC\cite{qsoc}.
  The  $j = 7/2$ states are near the Fermi level with bandwidth 0.092 eV, and $j = 5/2$ states are -1.309 eV below the Fermi level with bandwidth 0.086 eV.
 Both using GGA and GGA+SOC, the valence band maximum (VBM) is at $\Gamma$ point, and the conduction band minimum (CBM) at M point.  The GGA and GGA+SOC gap is 0.916 eV and 0.390 eV, respectively, and the gap reduce caused by SOC is 0.526 eV. The bands from -5.5 to -3.0 eV  are mainly
composed of I-p character states, slightly  hybridized with Yb-d/s ones.  The  hybridized Yb-d and I states are observed in the conduction bands.
When considering SOC, it is found that both conduction bands and valence bands below localized states move toward lower energy compared with ones using GGA.

\begin{figure}
    \includegraphics[width=8cm]{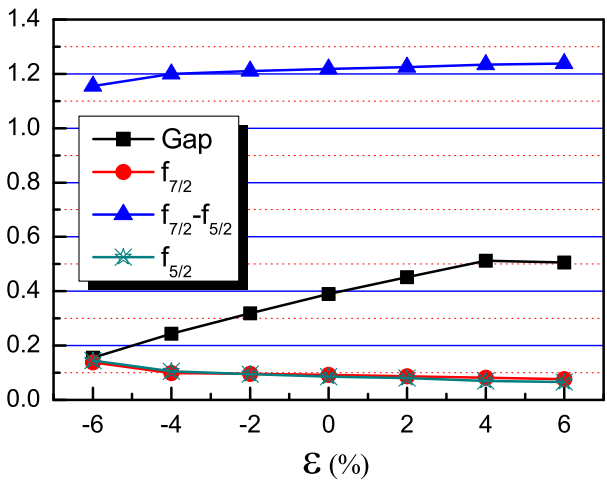}
  \caption{(Color online) The energy band gap ($Gap$), the gap between  the $j = 7/2$  and $j = 5/2$  states ($f_{7/2}$-$f_{5/2}$) and the widths of  the $j = 7/2$  and $j = 5/2$  states ($f_{7/2}$ and $f_{5/2}$) (unit: eV) as a function of $\varepsilon$.}\label{t5}
\end{figure}

The electronic transport coefficients of TMD monolayers have been widely investigated in theory\cite{q21,q22,q23}. It is natural to ask what effects on  transport coefficients can be induced by highly localized bands. Based on CSTA Boltzmann theory,  the Seebeck coefficient S  and electrical conductivity $\mathrm{\sigma/\tau}$ are performed   within rigid band approach.  It is noted that the calculated  $\mathrm{\sigma/\tau}$ depends  on $\tau$, while  S is independent of $\tau$.
 The n(p)-type doping effects can be  simulated by simply moving  the Fermi level  into conduction  (valence) bands, namely electron (hole) doping.
 The  room temperature S and  $\mathrm{\sigma/\tau}$ of $\mathrm{YbI_2}$  as  a function of doping level (N)   using GGA and GGA+SOC are plotted  in \autoref{t3}.  In low p-type doping, a detrimental influence on S caused by SOC   can be observed,  while  a neglectful effect on S (absolute value) in n-type or high p-type  doping can be found. For $\mathrm{\sigma/\tau}$, the SOC has little effect in both n- and p-type doping. It is clearly seen that the p-type S is larger than 300  $\mu$V/K  with doping level being up to 0.1 hole/per unit cell.  However, for n-type doping, the S can reach 300  $\mu$V/K, only below doping level of 0.008 electron/per unit cell. It is also found that the $\mathrm{\sigma/\tau}$ is very close to zero in p-type doping. These results are because the S
 is proportional to the effective mass of the carrier, while $\mathrm{\sigma}$ is inversely proportional to one.
 Therefore, highly localized bands can induce very large S and very low $\mathrm{\sigma}$ by producing very large effective mass of the carrier.
A similar effect can be found  in hole-doped PbTe or PbSe\cite{q24,q25}, and  the flat-band (localized bands) can be observed below their gaps.

During the fabrication process, 2D materials will commonly have residual strain. Next, we investigate the stability of highly localized bands again biaxial  strain.  The strain effects on the energy band structures and  transport properties of TMD monolayers have been widely investigated\cite{q22,q23,q26,q27}. The $\varepsilon=(a-a_0)/a_0$ is defined to simulate biaxial strain, in which $a_0$ is the unstrained lattice constant. The $\varepsilon$$<$($>$)0 means  compressive (tensile) strain. The related energy band structures of $\mathrm{YbI_2}$ monolayer are  shown  in \autoref{t4} using GGA+SOC, with strain from -6\% to 6\%.
In considered strain range, compressive strain can reduce the numbers of conduction band extrema (CBE) from two to one, while tensile strain can change relative position of two CBE. For example,  at 6\% strain, the CBM changes from M to $\Gamma$ point, which means that  tensile strain can induce conduction bands convergence between 4\% and 6\% strain, producing very large n-type S. Similar phenomenon caused by strain can also be found in TMD monolayers\cite{q22,q23,q27}.
The energy band gap, the gap between  the $j = 7/2$  and $j = 5/2$  states and the widths of  the $j = 7/2$  and $j = 5/2$  states  as a function of $\varepsilon$ are plotted in \autoref{t5}. With increasing strain, the energy band gap increases
from -6\% to 4\% strain, and then slightly reduces at 6\% strain. It is clearly seen that  the gap between  the $j = 7/2$  and $j = 5/2$  states and the widths of  the $j = 7/2$  and $j = 5/2$  states have very minor changes from -4\% to 6\% strain, and  the change is only 0.038 eV, 0.023 eV and 0.039 eV, respectively. So, the highly localized bands have very strong stability again strain.

\section{Discussions and Conclusion}
\begin{figure}
  \includegraphics[width=8cm]{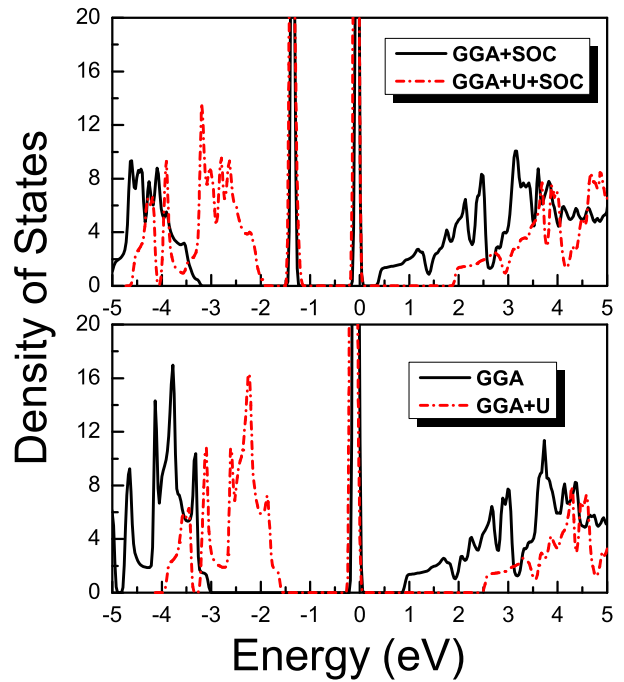}
\caption{The DOS  of $\mathrm{YbI_2}$ monolayer  using GGA+SOC, GGA+$U$+SOC, GGA and GGA+$U$. }\label{u}
\end{figure}

To account for $4f$ electron correlation effects, the onsite Coulomb interaction is included, and the  Coulomb potential $U_{eff}$  for the Yb-$4f$ orbitals is chosen to be 5 eV. The DOS  of $\mathrm{YbI_2}$  using GGA+SOC, GGA+$U$+SOC, GGA and GGA+$U$ are shown in \autoref{u}. Calculated results show that the isolated highly localized bands still exists, and only they  move to lower energies, which leads to increased energy band gap. When including onsite Coulomb interaction, the GGA gap changes from  0.92 eV to  2.53 eV, and GGA+SOC gap  from 0.39 eV to 1.94 eV. However, the gap between  the $j = 7/2$  and $j = 5/2$  states hardly changes. In fact, as $U$ is increased,  highly localized bands gradually move to lower energies. When $U$ is large enough, the $j = 5/2$  states  firstly cross with I-$p$ states, and then  the $j = 7/2$ states  also cross.

 In summary,  we investigate electronic structures  of  $\mathrm{YbI_2}$ monolayer, based mainly on the reliable first-principle calculations. Calculated results show that $\mathrm{YbI_2}$ monolayer is a indirect-gap semiconductor  using GGA+SOC, GGA+$U$+SOC, GGA and GGA+$U$, and the isolated highly localized bands with Yb-$4f$ character are observed in a very large gap of up to about 4 eV.   With the inclusion of SOC, the $4f$ bands are split into  the $j = 7/2$  and $j = 5/2$  states.   The isolated highly localized bands can induce very large effective mass of the  p-type carrier, and then leads to very large Seebeck coefficient and very low electrical conductivity in p-type doping.
Calculated results show that strain can tune the strength of conduction bands convergence by changing  relative position of  CBE. However, the strain has little effects on  isolated highly localized bands, namely they are stable again strain. The electron correlation effects only change the relative positions of  highly localized bands in the gap. Our works will motivate farther  experimental studies to synthesize $\mathrm{YbI_2}$ monolayer, and then to detect isolated highly localized bands.

\begin{acknowledgments}
This work is supported by the National Natural Science Foundation of China (Grant No. 11404391). We are grateful to the Advanced Analysis and Computation Center of CUMT for the award of CPU hours to accomplish this work.
\end{acknowledgments}


\begin{references}



\bibitem{q6}K. S.  Novoselov et al.,  Science \textbf{306}, 666	(2004).


\bibitem{q7}R. X.  Fei, W. B. Li, J. Li and L. Yang, Appl. Phys. Lett. \textbf{107}, 173104 (2015).


\bibitem{q8}J. P. Ji,   X. F. Song, J. Z. Liu et al.,  Nat. Commun. \textbf{7}, 13352 (2016).

\bibitem{q9}S. L. Zhang  M. Q. Xie, F. Y. Li, Z.  Yan, Y. F. Li, E. J. Kan,
W. Liu,  Z. F. Chen,  H. B. Zeng,  Angew. Chem. \textbf{128}, 1698 (2016).


\bibitem{q10}S. Balendhran, S. Walia, H. Nili, S. Sriram and M.Bhaskaran, small \textbf{11},  640 (2015).


\bibitem{q11}M. Chhowalla,	H. S. Shin,	G. Eda,	L. J.  Li,	K. P.  Loh	and H. Zhang, Nature Chemistry \textbf{5}, 263  (2013).


\bibitem{q12}B. Radisavljevic, A. Radenovic, J. Brivio, V. Giacometti and
A. Kis, Nat. Nanotechnol.  \textbf{6}, 147 (2011).

\bibitem{q13} D. Jariwala, V. K. Sangwan, L. J. Lauhon, T. J. Marks and
M. C. Hersam, ACS Nano  \textbf{8}, 1102 (2014).


\bibitem{q14}A. Y. Lu, H. Y. Zhu, J. Xiao et al., Nature Nanotechnology \textbf{12}, 744 (2017).



\bibitem{q15}H. Liu, A. T. Neal, Z. Zhu, Z. Luo, X. Xu, D. Tomnek, and
P. D. Ye, ACS Nano \textbf{8}, 4033 (2014).

\bibitem{q16}H. O. H. Churchill and P. Jarillo-Herrero, Nat. Nanotech. \textbf{9}, 330
(2014).

\bibitem{q17}B. J. Feng, B. T. Fu, S. Kasamatsu et al.,  Nat. Commun. \textbf{8}, 1007 (2017).

\bibitem{q18}C. Shao, X. X. Yu, N. Yang, Y. N. Yue  and H.  Bao, Nanosc. Microsc. Therm. \textbf{21}, 201 (2017).


\bibitem{q19}N. Mounet, M. Gibertini, P. Schwaller et al.,  Nat.  Nanotechnol. \textbf{13}, 246 (2018).

\bibitem{1}P. Hohenberg and W. Kohn, Phys. Rev. \textbf{136},
B864 (1964); W. Kohn and L. J. Sham, Phys. Rev. \textbf{140},
A1133 (1965).

\bibitem{2}P. Blaha, K. Schwarz, G. K. H. Madsen, D. Kvasnicka
 and J. Luitz, WIEN2k, an Augmented Plane Wave
+ Local Orbitals Program for Calculating Crystal Properties
(Karlheinz Schwarz Technische Universit\"at Wien, Austria) 2001,
ISBN 3-9501031-1-2


\bibitem{pbe}J. P. Perdew, K. Burke and M. Ernzerhof, Phys. Rev. Lett. \textbf{77}, 3865 (1996).

\bibitem{10}A. H. MacDonald, W. E. Pickett and D. D. Koelling, J. Phys. C \textbf{13}, 2675 (1980).

\bibitem{11}D. J. Singh and L. Nordstrom, Plane Waves, Pseudopotentials and the LAPW
Method, 2nd Edition (Springer, New York, 2006).

\bibitem{12}J. Kunes, P. Novak, R. Schmid, P. Blaha and
K. Schwarz, Phys. Rev. B \textbf{64}, 153102 (2001).

\bibitem{so}D. D. Koelling, B. N. Harmon, J. Phys. C: Solid State Phys.  \textbf{10}, 3107 (1977).



\bibitem{b}G. K. H. Madsen and D. J. Singh, Comput. Phys. Commun. \textbf{175}, 67
(2006).

\bibitem{q20}H. L. Zhuang and R. G. Hennig, J. Phys. Chem. C  \textbf{117}, 20440  (2013).

\bibitem{qsoc}T. R. Chang,  T. Das, P. J. Chen et al., Phys. Rev. B \textbf{91}, 155151 (2015).

\bibitem{q21}S. D. Guo and J. L. Wang, Semicond. Sci. Tech.
\textbf{31}, 095011 (2016).


\bibitem{q22}H. Y. Lv,   W. J. Lu,   D. F. Shao,  H. Y. Lub and   Y. P. Sun, J. Mater. Chem. C \textbf{4}, 4538 (2016).


\bibitem{q23}S. D. Guo, J. Mater. Chem. C  \textbf{4}, 9366 (2016).

\bibitem{q24}D. Parker and D. J. Singh, Phys. Rev. B \textbf{82}, 035204 (2010).

\bibitem{q25}D. J. Singh, Phys. Rev. B \textbf{81}, 195217 (2010).

\bibitem{q26}E. Scalise, M. Houssa, G. Pourtois, V. Afanas'ev  and A. Stesmans,   Nano Res. \textbf{5}, 43 (2012).

\bibitem{q27}D. Qin, X. J. Ge, G. Q. Ding, G. Y. Gao and J. T. Lv, RSC Adv. \textbf{7}, 47243 (2017).
\end{references}
\end{document}